\documentclass[showpacs,showkeys]{revtex4}
\usepackage{graphics,color}
\usepackage{amsmath,amssymb}

\begin{document}

\title{Stochastic Einstein equations}
\author{Vladimir Dzhunushaliev}
\email{vdzhunus@krsu.edu.kg}
\affiliation{
Institut f\"ur Physik, Universit\"at Oldenburg, Postfach 2503
D-26111 Oldenburg, Germany; \\
Institute for Basic Research, 
Eurasian National University, 
Astana, 010008, Kazakhstan; \\ 
Institute of Physicotechnical Problems and Material Science of the NAS
of the Kyrgyz Republic, 265 a, Chui Street, Bishkek, 720071,  Kyrgyz Republic 
}

\begin{abstract}
Stochastic Einstein equations are considered when 3D space metric
$\gamma_{ij}$ are stochastic functions. The probability density for the
stochastic quantities is connected with the Perelman's entropy functional. As an example, the Friedman Universe is considered. It is shown that for the Friedman Universe the dynamical evolution is not changed. The connection between general relativity and Ricci flows is discussed. 
\end{abstract}
\maketitle

\section{Introduction}

In quantum gravity dynamical variables are 3D space metric $\gamma_{ij}$ which
should be quantized, i.e. they have to be fluctuating quantities. Up to now we
do not have any quantum theory of gravity. Here we offer a stochastic approach
to quantum gravity where the dynamical variables are stochastic quantities
and they fluctuate according to some probability density. For the calculation
of this density we use the idea proposed in Ref. \cite{Dzhunushaliev:2008cz} that Ricci flow (for 3D space metric) is a statistical system and every metric in the Ricci flow is a microscopical state. 

There have been a number of investigations which aim to apply Ricci flow techniques in general relativity. In Ref.~\cite{Husain:2008rg} the evolution of wormhole geometries under Ricci flow is studied: depending on value of initial data parameters, wormhole throats either pinch off or evolve to a monotonically growing state. In Ref. \cite{Headrick:2006ti} Ricci flows are applied to general relativity, where gradient flow is Ricci flow, and the investigation is focused on the example of 4-dimensional Euclidean gravity with boundary $S^1 \times S^2$, representing the canonical ensemble for gravity in a box. In Ref. \cite{Samuel:2007ak} Ricci flow techniques is applied to general relativity: the evolution of the area ${\cal A}$ and Hawking mass ${\cal M}_H$ under the Ricci flow is studied. In Ref. \cite{Carfora:2008fr} Ricci flow deformation of cosmological initial data sets in general relativity is a technique for generating families of initial data sets which potentially would allow to interpolate between distinct spacetimes. The linear stability of Schwarzschild - Tangherlini spacetimes and their Anti-de Sitter counterparts under Ricci flow for a special class of perturbations in \cite{Dutta:2009pn} is investigated.

The main goal of this investigation is to show that there is some connection between 3D Ricci flow (for 3D space metric) and Einstein equations (for 4D metric). The connection is based on the idea that the 3D space metric in Einstein equations is a stochastic quantity with the probability density defined by the Ricci flow. 

\section{Stochastic Einstein equations}

According to Eq's \eqref{ham1} - \eqref{Kdot1} we will write the stochastic
Einstein equations in the form 
\begin{eqnarray}
 \tilde R - K_{ij}K^{ij} + K^2 &=& 2 \rho,
\label{2-10}\\
	D_j K^{j}_{\phantom{j}i} - D_i K &=& S_i,
\label{2-20}\\
	\frac{d}{dt} K_{ij} &=& - D_i D_j \alpha + \alpha ( \tilde R_{ij} 
        - 2 K_{il} K^l_{\phantom{l}j} + K K_{ij} - M_{ij} ) 
\label{2-30}
\end{eqnarray}
here $\tilde R_{ij}$ and $\tilde R$ are \textcolor{blue}{\emph{the fluctuating
3D Ricci tensor and scalar curvature}}. Eg's \eqref{2-10} - \eqref{2-30} are
stochastic equations (for details about stochastic equations, see appendix
\ref{sde}) since quantities $\tilde R_{ij}, \tilde R$ are made from a stochastic 
3D space metric $\tilde \gamma_{ij}$. The probability density $\rho$ for for the random functions $\tilde \gamma_{ij}$ is the functional $F(\lambda)$ which is connected with the Perelman's entropy functional $\mathcal W(\lambda)$ as 
$F(\lambda) = \frac{d \mathcal W(\lambda)}{d \lambda}$ (see, Appendix \ref{rflows} for the details). 

These equations describe the time evolution of the stochastic 3D space metric 
$\tilde \gamma_{ij}$. The next values of $\tilde \gamma_{ij}(t_0 + \Delta t)$ are defined through Eq's \eqref{2-10} - \eqref{2-30} taking into account the ``noise''  created by the stochastic terms $\tilde R_{ij}, \tilde R$. 

The solution procedure for Eq's \eqref{2-10} - \eqref{2-30} is following: 
\begin{enumerate}
 \item \label {proc1} the initial value of the 3D space metric 
 $\tilde \gamma_{ij}$ is chosen;
 \item the Ricci flow equations \eqref{4a-10} \eqref{4a-20} are solved in
	order to obtain the probability density for the ``noise'' 
	$\tilde \gamma_{ij}$;
 \item the equation \eqref{4a-40} is used for the definition of the probability
 	density for the ``noise'' $\tilde \gamma_{ij}, \tilde R_{ij}$ and $\tilde R$;
 \item the probability density \eqref{4a-40} is normalized;
 \item the probability density is used for obtaining of current values of the 
	stochastic quantities $\tilde \gamma_{ij}, \tilde R_{ij}$ and $\tilde R$ 
	that are inserted into Eq's \eqref{2-10} - \eqref{2-30}; 
 \item \label {proc2} Eq's \eqref{2-10} - \eqref{2-30} are used to define the
	metric 	$\tilde \gamma_{ij}(t_0 + \Delta t)$;
 \item the steps \ref{proc1} - \ref{proc2} are being repeated to obtain 
	$\tilde \gamma_{ij}(t_0 + n \Delta t)$. 
\end{enumerate}

\section{Stochastic Friedman Universe}

Now we would like to apply Eq's \eqref{2-10} - \eqref{2-30} to describe the 
Friedman Universe where the 3D space metric is a stochastic quantity. The Einstein
equations for the Friedman Universe with metric 
\begin{equation}
	ds^2 = dt^2 - a^2(t) \left[ 
		d \chi^2 + \sin^2 \chi \left(  
			d \theta^2 + \sin^2 \theta d \phi^2
		\right) 
	\right] 
\label{3-10}
\end{equation}
are 
\begin{eqnarray}
 	3 \frac{\dot a^2}{a^2} + \frac{3}{a^2} &=& \varkappa \varepsilon,
\label{3-20}\\
	\frac{\ddot a}{a} + \frac{\dot a^2}{a^2} + \frac{1}{a^2} &=& 0
\label{3-30}
\end{eqnarray}
here, for simplicity, we consider dust matter with zero pressure $p=0$.
We use equation \eqref{3-20} for the determination of the energy density
$\varepsilon$, and equation \eqref{3-30} is the stochastic equation for the
determination of the scale factor $a(t)$. 

First we will seek for the Ricci flow equations \eqref{4a-10} \eqref{4a-20}.
The 3D space metric of the Friedman Universe is 
\begin{equation}
	dl^2 = a^2(t) \left[ 
		d \chi^2 + \sin^2 \chi \left(  
			d \theta^2 + \sin^2 \theta d \phi^2
		\right) 
	\right] .
\label{3-40}
\end{equation}
In this case the Ricci flow equations \eqref{4a-10} \eqref{4a-20} are 
\begin{eqnarray}
 	\frac{d a^2}{d \lambda} &=& -4 ,
\label{3-50}\\
	\frac{d f}{d \lambda} &=& - \frac{6}{a^2} + 
	\frac{3}{2} \frac{1}{\lambda_0 - \lambda}
\label{3-60}
\end{eqnarray}
where $\lambda$ is the parameter under which the Ricci flow is evolved. The
solution of the system \eqref{3-50} \eqref{3-60} is 
\begin{eqnarray}
 	a &=& 2 \sqrt{\left( \lambda_0 - \lambda \right) } ,
\label{3-70}\\
	f &=& f_0 = \mathrm{const}.
\label{3-80}
\end{eqnarray}
Then the functional \eqref{4a-40} is equal to zero 
\begin{equation}
	F(\lambda) = 0. 
\label{3-90}
\end{equation}
We choose the probability density for the fluctuating 3D space metric 
$\tilde \gamma_{ij}$ in the stochastic Einstein equations as the functional \eqref{4a-40}. This means that for every moment $t$ the probability density for the ``noise'' 
$\tilde g_{ij}, \tilde R_{ij}, \tilde R$ is the functional \eqref{4a-40}. Thus, for solving stochastic Einstein equations \eqref{3-30} we have to use a randomly function $\tilde g_{ij}$ with the probability density $\rho$ equal to the functional 
$\rho (\lambda) = F(\lambda) = \frac{\mathcal W(\lambda)}{d \lambda}$. In order
to have the full probability equal to unity we choose 
\begin{equation}
	\int \rho dV = 1 
\label{3-100}
\end{equation}
and complement the function $\rho(\lambda)$ by delta - function at 
$\lambda = 0$ 
\begin{equation}
	\rho(\lambda) = \delta(\lambda) .
\label{3-110}
\end{equation}
Then 
\begin{equation}
	\rho(\lambda) = \begin{cases} 
	\mathcal W(\lambda),  & \text{ if } \lambda \neq 0 
	\text{ that is consistent with equation \eqref{3-90}}\\
	\delta(\lambda), & \text{ if } \lambda = 0 
	\end{cases}
\label{3-120}
\end{equation}
The probability distribution \eqref{3-120} means that at any moment $t$ the
stochastic function $\tilde a(t)$ is realized with the probability equal to
unity, i.e. the Friedman equations \eqref{3-20} - \eqref{3-30} do not change. The
3D metric $a(t)$ is the determinate function. One can say that in this sense 4D Einstein 
equations and 3D Ricci flows are connected in the manner discussed above. 

\section{Discussion and conclusions}

Here we have offered the idea that the Ricci flows can be applied in general relativity. For every moment the Ricci flow shows a probability for the realization of 3D space metric in the Einstein equations. In such approach the Einstein equations become stochastic equations where dynamical variables (3D space metric) are stochastic functions. The probability density of these stochastic variables is connected with some functional of the Ricci flow of the 3D space metric. In this sense the 4D Einstein equations can be connected with the 3D Ricci flows. 

Naturally the question arises: whether such an approach has the connection with quantum gravity ? At the moment we cannot give the answer on this question but there are some arguments that such connection may exist. It is well known that in quantum theory the path integral can be approximately calculated if a classical solution does exist (saddle-point method). One can assume that in  some way the calculation of a path integral can be connected with Ricci flow.

Here we have shown that in a special case of the Friedman Universe the dynamics of the scale factor have not changed. It is interesting to do similar calculations for another solutions of the Einstein equations: black holes, wormholes, cosmological solutions and so on. 

\section*{Acknowledgements}

I am grateful to the Research Group Linkage Programme of the Alexander  von
Humboldt Foundation for the support of this research and would like to express
the gratitude to the Department of Physics of the Carl von Ossietzky University
of Oldenburg  and, specially, to J. Kunz. 
\appendix

\section{3+1 - decomposition}

In this section we follow to Ref. \cite{Baumgarte:1998te}. We write the metric
in the form
\begin{equation}
	ds^2 = - \alpha^2 dt^2 + \gamma_{ij} (dx^i + \beta^i dt)(dx^j + \beta^j dt),
\label{1a-10}
\end{equation}
where $\alpha$ is the lapse function, $\beta^i$ is the shift vector, and
$\gamma_{ij}$ is 3D spatial metric; Latin indexes are spatial indexes $i,j =
1,2,3$; Greek indexes are spacetime indexes $\mu, \nu =0,1,2,3$.  The extrinsic
curvature $K_{ij}$ can be defined by the equation
\begin{eqnarray} 
\label{gdot1}
	\frac{d}{dt} \gamma_{ij} &=& - 2 \alpha K_{ij},
\\
	\frac{d}{dt} &=& \frac{\partial}{\partial t} - {\cal L}_{\beta}
\end{eqnarray}
where ${\cal L}_{\beta}$ denotes the Lie derivative with respect to
$\beta^i$.

Einstein equations 
\begin{equation}
	\mathcal R_{\mu \nu} - \frac{1}{2} g_{\mu \nu} \mathcal R = 
	T_{\mu \nu} 
\label{1a-20}
\end{equation}
can then be splitted into the Hamiltonian constraint
\begin{equation} 
\label{ham1}
	R - K_{ij}K^{ij} + K^2 = 2 \rho,
\end{equation}
the momentum constraint
\begin{equation} 
\label{mom1}
	D_j K^{j}_{\phantom{j}i} - D_i K = S_i,
\end{equation}
and the evolution equation for the extrinsic curvature
\begin{equation} 
\label{Kdot1}
	\frac{d}{dt} K_{ij} = - D_i D_j \alpha + \alpha ( R_{ij} 
        - 2 K_{il} K^l_{\phantom{l}j} + K K_{ij} - M_{ij} ) 
\end{equation}
here $D_i$ is the covariant derivative associated with $\gamma_{ij}$; 
$\mathcal R_{\mu \nu}$ is the 4D Ricci tensor; $\mathcal R$ is the 4D scalar
curvature; $R_{ij}$ is the three-dimensional Ricci tensor
\begin{eqnarray} 
\label{ricci}
	R_{ij} & = & \frac{1}{2} \gamma^{kl} 
	\Big( \gamma_{kj,il} + \gamma_{il,kj} 
		- \gamma_{kl,ij} - \gamma_{ij,kl} \Big) \\
	[1mm]
 	& & + \gamma^{kl} \Big( \Gamma^m_{il} \Gamma_{mkj}
 	- \Gamma^m_{ij} \Gamma_{mkl} \Big), \nonumber
\end{eqnarray}
and $R$ is its trace of $R = \gamma^{ij} R_{ij}$.  We have also introduced the
matter sources $\rho$, $S_i$ and $S_{ij}$, which are projections of
the stress-energy tensor with respect to the unit normal vector
$n_{\alpha}$,
\begin{eqnarray}
	\rho & = & n_{\alpha} n_{\beta} T^{\alpha \beta}, \nonumber \\[1mm]
	S_i  & = & - \gamma_{i\alpha} n_{\beta} T^{\alpha \beta}, \\[1mm]
	S_{ij} & = & \gamma_{i \alpha} \gamma_{j \beta} T^{\alpha \beta}, \nonumber
\end{eqnarray}
and have abbreviated
\begin{equation}
	M_{ij} \equiv S_{ij} + \frac{1}{2} \gamma_{ij}(\rho - S),
\end{equation}
where $S$ is the trace of $S_{ij}$, $S = \gamma^{ij} S_{ij}$.

The evolution equations~(\ref{gdot1}) and~(\ref{Kdot1}) together with
the constraint equations~(\ref{ham1}) and~(\ref{mom1}) are equivalent
to the Einstein equations, and are commonly referred to as the ADM
form of the gravitational field equations~\cite{adm62}.  

\section{Stochastic equations}
\label{sde}

The definition of a \emph{stochastic differential equation} is as follows
(for details, see Ref. \cite{grispin})  
\begin{eqnarray}
 d \textbf{X}(t) &=& \textbf{b} \left( \textbf{X}(t) \right) dt + 
	\textbf{B} \left( \textbf{X}(t) \right) d \textbf{W}(t), 
\label{3a-10}\\
	\textbf{X}(0) &=& x_0.
\label{3a-20}
\end{eqnarray}
where $\textbf{b}$ is a given vector field; $\textbf{X}$ is a solution
which is random field and $\textbf{W}$ is a stochastic ''noise''. 

\section{Ricci flows}
\label{rflows}

\emph{Ricci flow} on a 3D space with positive metric is defined as follows (for
details, see Ref. \cite{topping}) 
\begin{eqnarray}
	\frac{\partial \gamma_{ij}}{\partial \lambda} &=& -2 R_{ij}, 
\label{4a-10}\\ 
	\frac{\partial f}{\partial \lambda} &=& -\Delta f + 
	\left |	\nabla f \right |^2 - R + 
	\frac{3}{2 } \frac{1}{\left( \lambda_0 - \lambda \right)}
\label{4a-20}
\end{eqnarray}
where $\lambda$ is the parameter describing the evolution of the metric $\gamma_{ij}$
under Ricci flow and $\lambda_0$ is the value of the parameter $\lambda$ where
$\gamma_{ij}$ becomes singular. For the Ricci flow one can introduce the
following functionals
\begin{eqnarray}
	\mathcal W(\lambda) &=& \int \left[ 
		\left( \lambda_0 - \lambda \right)
	\left( 
			R + \left | \nabla f \right|^2
		\right) + f -3 
 	\right] u dV, 
\label{4a-30}\\ 
	F(\lambda) &=& \frac{d \mathcal W}{d \lambda} = 
	2 \left( \lambda_0 - \lambda \right)
	\int \left[ R_{ij} + \frac{\partial^2 f}{\partial x^i \partial x^j} 
	- \frac{\gamma_{ij}}{2 \left( \lambda_0 - \lambda \right)}
	\right] 
	\left[ R^{ij} + \frac{\partial^2 f}{\partial x_i \partial x_j} 
	- \frac{\gamma^{ij}}{2 \left( \lambda_0 - \lambda \right)}
	\right] 
	u dV \geq 0 ,
\label{4a-40}\\
	u &=& \frac{1}{\left( 4 \pi \right)^{3/2}} 
	\frac{1}{\left( \lambda_0 - \lambda \right)^{3/2}} 
	e^{-f}
\label{4a-50}
\end{eqnarray}
where $\mathcal W(\lambda)$ is the Perelman's entropy functional.

\end{document}